\documentclass[prl,reprint,aps]{revtex4-1}

\usepackage{graphicx}
\begin{document}
%

\title{Separating Pathways in Double-Quantum Optical Spectroscopy Reveals Excitonic Interactions}
%
\author{Jonathan O. Tollerud, and Jeffrey A. Davis*}
%
%
\affiliation{%
	Centre for Quantum and Optical Science, Swinburne University of Technology, Hawthorn, VIC, Australia
}
\begin{abstract}Techniques for coherent multidimensional optical spectroscopy have been developed and utilised to understand many different processes, including energy transfer in photosynthesis and many-body effects in semiconductor nanostructures. Double-quantum 2D spectroscopy is one variation that has been particularly useful for understanding many-body effects. 
	In condensed matter systems, however, there are often many competing signal pathways, which can make it difficult to isolate different contributions and retrieve quantitative information. 
	Here, a means of separating overlapping pathways while maintaining the fidelity of the relevant peak/s is demonstrated. 
	This selective approach is used to isolate the double-quantum signal from a mixed two exciton state in a semiconductor quantum well. The removal of overlapping peaks allows analysis of the relevant peak-shape and thus details of interactions with the environment and other carriers to be revealed. An alternative pulse ordering identifies a double-quantum state associated only with GaAs defects, the signature of which has previously been confused with other interaction induced effects. The experimental approach described here provides access to otherwise hidden details of excitonic interactions 
	and demonstrates that the manner in which the double-quantum coherence is generated can be important and provide an additional control to help understand the many-body physics in complex systems.
\end{abstract}

\email{JDavis@swin.edu.au}
%
\keywords{Coherent Multidimensional Spectroscopy, Ultrafast Dynamics, Semiconductor nanostructures, Excitons, Many-body Effects}
%
\maketitle

\section{Introduction}

Techniques for multidimensional nuclear magnetic resonance spectroscopy (NMR) have been developed over the course of several decades to the point where modern pulse sequences may contain hundreds of precisely controlled pulses and can be used to determine complex protein structures~\cite{Wuthrich2003}. Coherent multidimensional electronic spectroscopy (CMDS)\cite{Hybl2001} is conceptually similar to multidimensional NMR. Indeed, several concepts developed in NMR spectroscopy (e.g. phase cycling~\cite{Bodenhausen1977,Vaughan2007,Tian2003}, rotating frame detection, and double-quantum spectroscopy~\cite{Vega1976}) have been applied to experiments at optical frequencies, while several others have been considered~\cite{Mukamel2000,Mukamel2009a}.

Optical CMDS has found application across many material systems where it has been used to reveal the role of many-body effects in semiconductor nanostructures~\cite{Li2006,Karaiskaj2010,Turner2012,Nardin2014}, quantum effects in biological light harvesting complexes~\cite{Engel2007,Panitchayangkoon2010,Westenhoff2012,Collini2010a,Romero2014}, radiative lifetimes in monolayer semiconductors~\cite{Moody2015} and to directly track multi-step energy transfer~\cite{Zhang2015}, for example. The majority of these applications have utilised the 3rd order response to generate 2D spectra based on rephasing or non-rephasing scans and occasionally double-quantum (2Q) 2D scans. Within these experiments the aims have been to resolve and understand individual states or the interactions between pairs of states. 

Higher order experiments have also begun to find some application, for example, Zhang et al. \cite{Zhang2015} have utilized a 5th order experiment to track multi-step energy transfer processes in LHC-II. The limitation with the approach they use is that to obtain the full 5 dimensional data set (or even a 3D dataset) takes many days. The approach adopted in NMR is to tailor the pulses to select specific pathways so that full scans are not required. This is referred to as heteronuclear NMR, because the different pulses can be resonant with different nuclear spin transitions. This type of approach has been demonstrated for 3rd order optical experiments~\cite{Tollerud2014,Senlik2015,Yuen-Zhou2014a}, where the excitation pulses are spectrally shaped to select specific states, and could in principle be extended to the 5th order experiments.
Higher-order experiments can also be used to excite multi-quantum coherences, which can provide details of the interactions between the excited systems. In NMR this approach is used to select regions of the molecule where two, three or more nuclei interact with one another in-turn. This type of filtering enables the identification and selection of sequences of nuclei in complex molecules and subsequently the neighbouring or nearby atoms and sequences, and ultimately the complex molecular structure. Optical 2Q spectroscopy has been used to explore 2-exciton correlations in semiconductor nanostructures~\cite{Tollerud2016a,Stone2009,Karaiskaj2010} and higher electronic states in molecular systems\cite{Christensson2010}. 

Higher order correlations (e.g. 3Q and 4Q) have also been explored with 5th and 7th order experiments to understand interactions between 3 and 4 excitons in semiconductor quantum wells~\cite{Bolton2000,Turner2010b}. This type of high-order correlation spectroscopy has the potential to provide insight into many of the material systems that are receiving a great deal of attention because of the many-body effects and correlated carrier interactions~\cite{Chemla2001,Li2006,Nardin2014}, including topological insulators and high temperature superconductors. 

The typical approach to 2Q experiments (and the equivalent is true for higher order experiments) is to use two broadband pulses, coincident in time, to excite the 2Q coherences, which are then allowed to evolve before a subsequent pulse drives the system to a 1-quantum coherence that radiates the detected signal. The 2Q coherence can involve identical excitons or excitons that are different, but which are in some way coupled (resulting in a `mixed 2-exciton state'). In the case of a simple two-body picture, it doesn't matter which exciton in the two-exciton state is excited first, all that matters is the evolution of the 2Q coherence and final 1Q coherence. In systems where interactions and many-body effects are important, however (i.e. most solid state materials), the order of interaction can be important, and furthermore, the time spent in the single exciton coherence can affect the 2Q coherence.

Here we combine the ability to shape the spectral amplitude of the excitation pulses \cite{Tollerud2014} with the introduction of delays between the pulses generating the 2Q coherence to isolate different pathways and demonstrate that in semiconductor quantum wells, the order and timing of excitation does matter. Furthermore, using this approach it becomes possible to remove contributions from free-carriers and reveal the natural lineshape of the mixed 2-exciton coherence. With the same set of data we identify a doubly excited state associated with GaAs defects, the nature of which only becomes apparent when the order of interactions is considered.

\begin{figure}[t!]
	\centering
	\includegraphics[width=\linewidth]{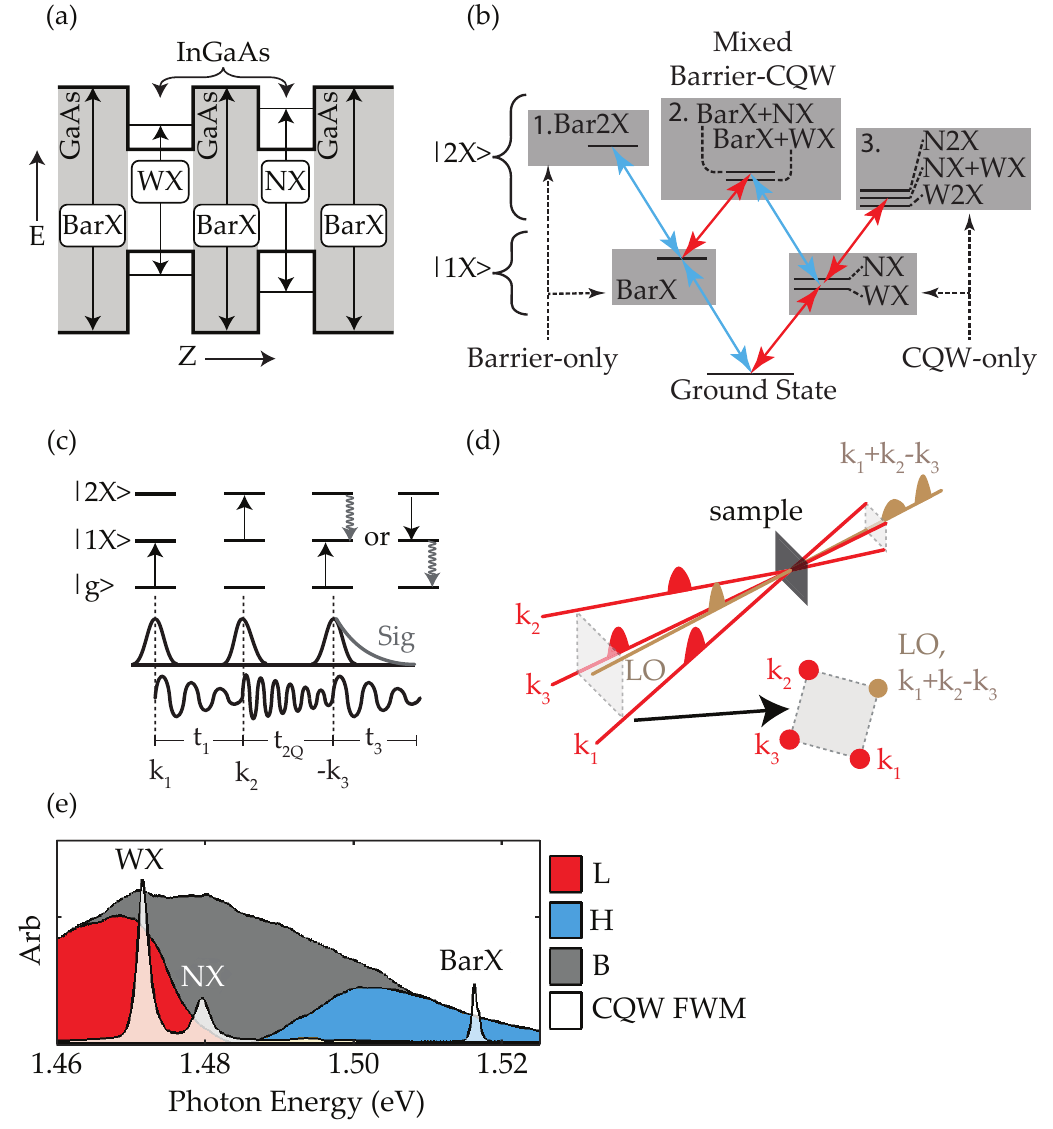}
	\caption{\label{FigSample} (a) A cartoon of the InGaAs/GaAs coupled quantum well sample used in this experiment. (b) the 10 level system from the three excitonic states including all two-exciton states. Grouped for convenience into a quasi-six level system. (c) The pulse ordering used in double-quantum experiments and the state of the system during each time period along with the effect of each subsequent laser pulse. (d) the square beam geometry used in this double-quantum experiment showing the three excitation beams ($k_1$, $k_2$, and $k_3$) and the reference (LO) (e) the excitation spectra and InGaAs FWM signal at $t_3=t_{2Q}=0$\,fs.%
	}
\end{figure}

\section{Coupled quantum well sample}

The coupled-quantum well (CQW) sample we study  consists of two In$_{0.05}$Ga$_{0.95}$As QWs with GaAs barriers, as depicted in Fig.\ref{FigSample} (a). The two QWs have slightly different widths (8\,nm and 10\,nm) so that the exciton transitions can be separated spectrally. The wells are separated by a 10\,nm central barrier. Previously, this sample has been used to investigate the nature of inter-well interactions~\cite{Nardin2014}, the interaction between `dark' and bright states~\cite{Tollerud2016b}, and the dynamics of the QW excitons at very low excitation density~\cite{Tollerud2016a}, but here we focus exclusively on interactions between excitons in the barrier and excitons in the QWs. For the experiments reported here, the sample was kept at 6\,K in a vibration isolated recirculating helium cryostat.

There are three `bright' exciton transitions in this sample - one predominately localized in each of the QWs (WX and NX), and one in the barrier (BarX). As described in more detail in the supporting information, this sample can be represented as a 10-level system (10LS), including three one-exciton and six two-exciton states, which can be grouped into the six types shown in Fig.~\ref{FigSample} (b). 
The two-exciton states are separated into three groups: Box 1 (3) contains all two exciton states in which both excitons are in the barrier (CQW). Box 2 includes the mixed two exciton states in which one exciton is in the barrier and the other is in a QW. The selective pulse sequences used in this paper employ spectral amplitude masks and pulse orderings which were chosen to isolate signals involving the two-exciton states in box 2 and eliminate those involving the states in boxes 1 and 3. 

\section{Experimental}
\subsection{Double-quantum spectroscopy} 

Interactions and correlations among excitons (electron-hole pairs) in semiconductor QWs can be studied using double-quantum 2D Fourier transform spectroscopy~\cite{Tollerud2016a,Turner2009,Turner2010b,Karaiskaj2010}. In this type of experiment, two ultrafast pulses excite a coherent superposition of the ground state (g) and a doubly excited state (2), as shown in Fig. \ref{FigSample} (c). In such a system, the g$\leftrightarrow$2 transition is forbidden, but a coherent superposition of g and 2 can be generated through a two-step process. The first pulse (resonant with the g$\leftrightarrow$1 transition) excites a coherent superposition of the ground state and a singly excited state. The second pulse (resonant with the 1$\leftrightarrow$2 transition, which may or may not be the same as the g$\leftrightarrow $1 transition) converts this into a coherent superposition of the ground state and the doubly excited state. 
A third pulse then interrogates the system by converting the coherent superposition of g and 2 into a coherent superposition of g and 1 or 1 and 2, which relaxes to state g or 1, respectively, emitting the signal photons in the process. The spectral phase and amplitude of the signal ($E_3$) are recorded as a function of the time delay between the second and third pulses ($t_{2Q}$). A Fourier transform as a function of $t_{2Q}$ allows us to display the recorded data as a double-quantum 2D spectrum, in which the complex signal is plotted as a function of the energy of the g - 2 coherence, $E_{2Q}$, and the energy of the emitted signal, $E_3$.

In semiconductor nanostructures, the double-quantum state probed with double-quantum spectroscopy is typically a two-exciton state. If both excitons are identical, then the signal will appear as a peak along the $E_{2Q} =2 E_{3}$ line (from here referred to as the diagonal line) since the $g\leftrightarrow 1$ and $1\leftrightarrow 2$ transition energies are equal. Such peaks are typically referred to as diagonal peaks (DPs). Any shifts of the $1\leftrightarrow 2$ transition energy from the $g\leftrightarrow 1$ transition energy due to, for example the formation of a bound biexciton, will shift the peak off the diagonal. Similarly, if the two-exciton state is made up of two different types of excitons (from here called a mixed two-exciton state), then the $g\leftrightarrow 1$ and $1\leftrightarrow 2$ transition energies are no longer equal and the signal will appear as a peak away from the diagonal line (referred to as cross-peaks or CPs). 

Signals in this type of double-quantum experiments only arise when there is some interaction or correlation between the pairs of excitons that make up the two-exciton state\cite{Mukamel2007,Gellen}. The position and shape of these peaks can then provide details of the interactions. Isolating the peaks corresponding to specific well-defined pathways, however, is not always easily accomplished. There can be - and frequently are - many overlapping contributions that can be difficult to separate. Furthermore, additional carriers excited into the system but not directly involved in generating the signal can alter the peak shapes, and in some cases the position and amplitude of the peaks as well~\cite{Tollerud2016a}. In a many body system where correlations are important, how you get to the two-exciton state can become important. For example, whether you get to the AB two-exciton state via state A or via state B can affect the ensuing 2D spectrum, even though in a simple non-interacting system this should not be the case.
To fully understand the interactions it then becomes necessary to isolate the specific quantum pathway of interest including the excitation order and minimise the interactions with additional carriers. This selectivity can be achieved by spectrally shaping the excitation pulses and carefully controlling the delay between the first two pulses.

\subsection{Pulse-shaper based selective double-quantum spectroscopy}

To perform the selective double-quantum measurements, we use a spatial light modulator (SLM) based multidimensional spectroscopy apparatus similar to the one originally developed by the Nelson group at MIT~\cite{Turner2011,Tollerud2014}. The four beams required for double-quantum 2D spectroscopy are generated in the rotated box geometry shown in Fig.~\ref{FigSample} (d) by an SLM based Fourier beam-shaper. A pulse shaper based on a phase only SLM is then used to individually control the inter-pulse delays and compress the pulses with linear and nonlinear spectral phase patterns (respectively). A vertical sawtooth phase grating is applied along the vertical axis of the SLM so that the shaped beams are diffracted down and can be picked off. The spectral amplitude of the excitation pulses is controlled by varying the depth of the vertical sawtooth phase grating, as demonstrated previously~\cite{Tollerud2014,Vaughan2007}. This allows independent control of the spectral amplitude and phase distributions of each of the four beams, which enables selective two-quantum experiments. Changing the pulse sequence, spectra or phase is then as simple as changing the phase patterns sent to the SLM. 

In this experimental configuration all of the beams are incident upon the same optics and so remain intrinsically phase-locked (better than $\lambda$/1000 over 30 minutes, see supporting information) a key requirement and often a challenge in double-quantum spectroscopy. The average photon density for each focused pulse was $3.4\times10^{10}$cm$^{-2}$ ($2.2\times10^{10}$cm$^{-2}$) for the broadband (shaped) pulses. The beams were co-linearly polarized and the delay between the first two pulses (t$_1$) was set to 300\,fs for all measurements described here. The pulse durations were measured using cross-correlations, and found to be $\sim$48\,fs, 84\,fs, and 115\,fs for the broadband pulse, shaped low energy pulse (L) that is resonant with only the CQW states, and shaped high energy pulse (H) that is resonant with only the Barrier states, respectively. More experimental details are shown in the supporting information.

\begin{figure}[t!]
	\centering
	\includegraphics[width=\linewidth]{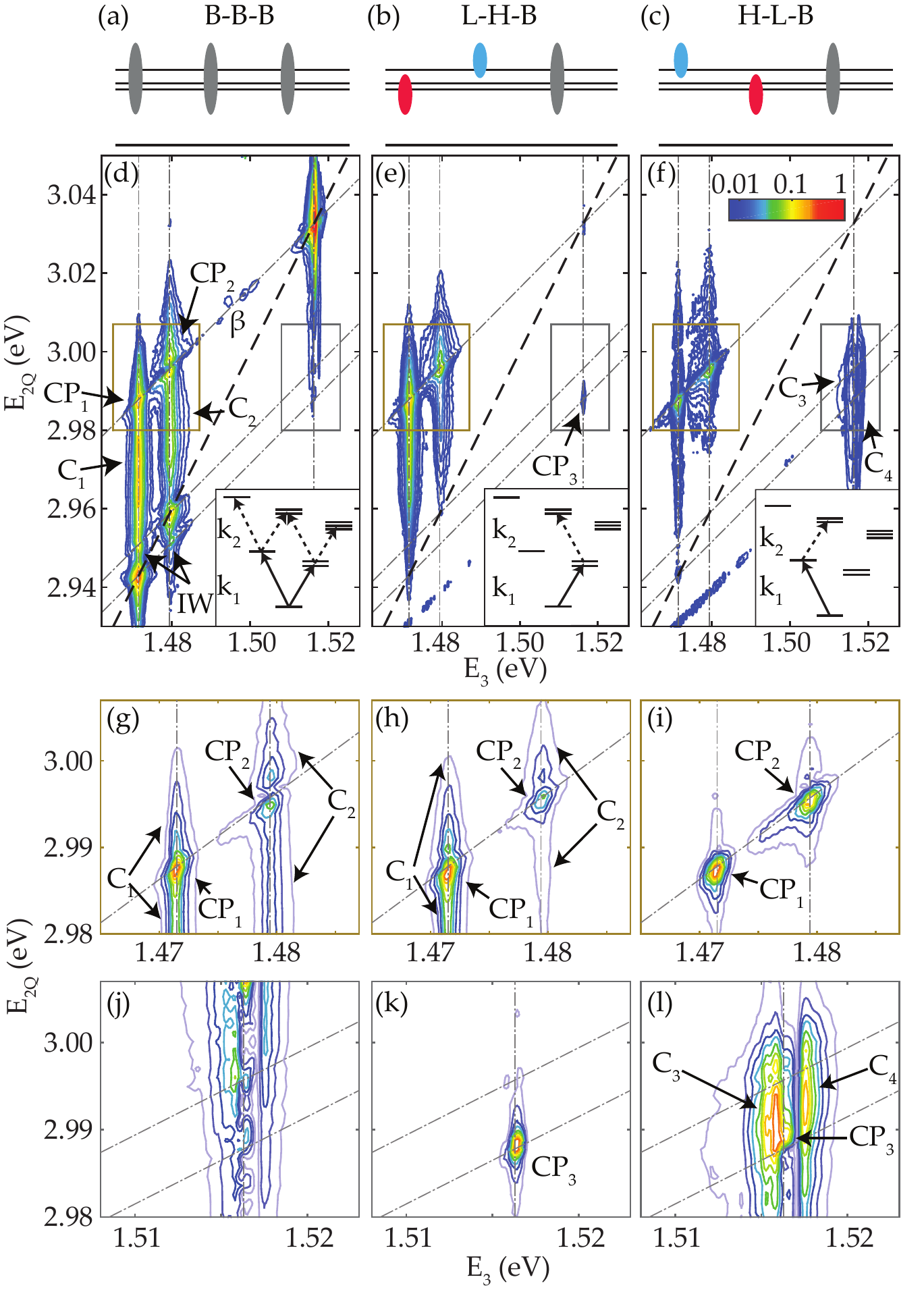}
	\caption{\label{SpectrumFull} (a)-(c) B-B-B, L-H-B and H-L-B sequences (respectively). (d)-(f) Double-quantum spectra for B-B-B, L-H-B and H-L-B sequences (respectively). The insets show the pathways that are possible for each sequence. (g)-(i) show a zoomed in view of the above-diagonal CPs. (j)-(l) shows a zoomed in view of the below-diagonal CPs.%
	}
\end{figure}

\section{Results}

Double-quantum 2D spectra were recorded using three pulse sequences: B-B-B, a standard double-quantum pulse sequence with three identical broadband pulses resonant with all the transitions in Fig~\ref{FigSample} (e); L-H-B, a pathway selective pulse sequence that excites the mixed two-exciton states in Box 2 of Fig~\ref{FigSample} (b), with the first pulse resonant with the WX and NX, and the second pulse resonant with BarX; and H-L-B, similar to the L-H-B sequence but with the first two pulses reversed such that the first pulse was resonant with BarX and the second pulse resonant with WX and NX.
The three pulse sequences are also depicted graphically in Fig.~\ref{SpectrumFull} (a)-(c), and the available pathways up to the various two-exciton states are shown in the insets of Fig.~\ref{SpectrumFull} (d)-(f). A full list of the available pathways is provided using two-sided Feynman-Liouville diagrams in the supporting information.

Fig.~\ref{SpectrumFull} (d)-(f) show double-quantum absolute value spectra collected using the B-B-B, L-H-B, and H-L-B pulse sequences, respectively. 
The B-B-B spectrum (Fig.~\ref{SpectrumFull} (d) ) shows three DPs (WX, NX, BarX) and several CPs. 
A close up of the above-diagonal CPs labelled CP$_1$ and CP$_2$ is shown in Fig.~\ref{SpectrumFull} (g). CP$_1$ and CP$_2$ correspond to the mixed two-exciton coherence, with E$_3$ equal to WX and NX, and E$_{2Q}$ equal to the WX+BarX and NX+BarX mixed two-exciton energies, respectively. These two peaks overlap signals that stretch over more than 40\,meV along the E$_{2Q}$ axis. These continuum signals, labeled C$_1$ and C$_2$ correspond to double-quantum coherences that involve one exciton in a QW (WX and NX, respectively) and one unbound electron-hole pair (EHP) in either the QW, or in the barrier. These exciton-EHP double-quantum coherences dephase rapidly as a function of t$_{2Q}$, and hence are broad along E$_{2Q}$.  

The corresponding below-diagonal CPs involving the same mixed two-quantum coherences but emitting with the energy of BarX are missing, or hidden by other contributions and/or interaction induced effects. A close-up of the region where these CPs are expected, is shown in Fig.~\ref{SpectrumFull} (j). There may be peaks from the mixed two-exciton coherence, but due to the significant broad continuum peaks these are not clear enough to definitively identify. 

We also observe NX-WX inter-well CPs (labelled `IW') as well as CPs for interactions of WX with a dark excitonic state (labelled `$\beta$') which appears with emission at 1.493\,eV. The $\beta$ exciton state is a spatially indirect exciton consisting of an electron in the wide well and a hole in the barrier, and is described in detail elsewhere~\cite{Tollerud2016b}. In this article, our focus is on WX-BarX and NX-BarX mixed two-exciton states.

The spectrally shaped pulse sequences (L-H-B and H-L-B) lead to significant changes in the 2Q spectra. First, in both Fig.~\ref{SpectrumFull} (e) and Fig.~\ref{SpectrumFull} (f) the DPs and IW CPs are no longer present, whereas the CP$_1$ and CP$_2$ above-diagonal CPs remain and one of the expected below-diagonal CPs (CP$_3$) becomes evident. The details of these CPs can be seen more clearly in Fig.~\ref{SpectrumFull} (g-l) which plot these regions on a linear colour scale. 
CP$_3$ (the below-diagonal equivalent of CP$_1$) is clearly evident in  Fig.~\ref{SpectrumFull} (k), (L-H-B sequence)  and 
can also be identified in Fig.~\ref{SpectrumFull} (l) (H-L-B sequence), although there the excitonic peak is overlapped with two strong continuum features (C$_3$ and C$_4$). C$_3$ and C$_4$ have emission energies below and above BarX, respectively, and correspond to known emission energies of GaAs defects~\cite{Gilliland1997,Skolnick1986,Alessi1997}.  It is clear from these plots that the spectra measured 
are significantly different for the two selective pulse sequences, suggesting that the order of the interactions does indeed matter in these CQW samples.

\begin{figure}[htbp]
	\centering
	\includegraphics[width=\linewidth]{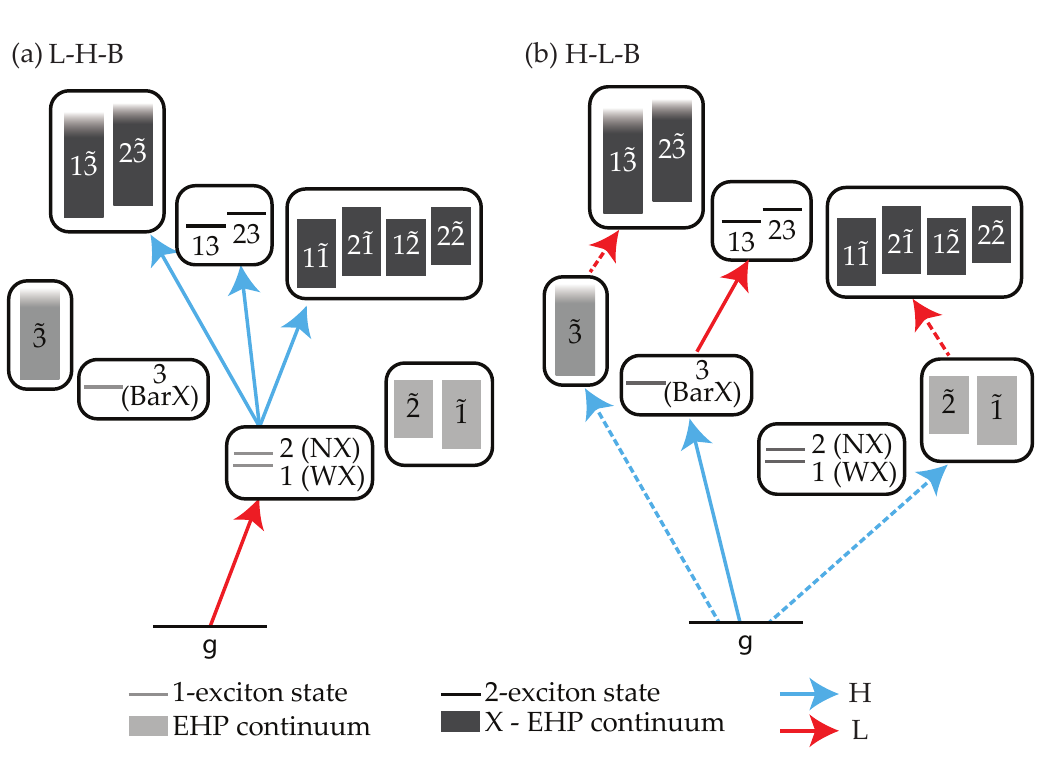}
	\caption{The states excited by (a) L-H-B and (b) H-L-B sequences show how the 2Q coherences involving unbound electron-hole pairs can be excited in the L-H-B pulse sequence, but in H-L-B pulse sequence can be eliminated by introducing a delay between the first two pulses to ensure the 1Q free carrier coherence decays before the second pulse arrives (dashed arrows). The WX, NX and BarX labels are replaced with labels of 1, 2 and 3, respectively, for the excitons, $\tilde{1}$, $\tilde{2}$, and $\tilde{3}$ for the EHP localised in the wide well, narrow well and barrier, respectively. 
		\label{2Qpathways}%
	}
\end{figure}

\section{Discussion}

In order to understand the differences observed for the two pathway selective sequences, consider first the above-diagonal peaks in Fig. \ref{SpectrumFull} (k-l). Double-quantum coherences that involve one QW exciton and an unbound electron-hole pair (in the QW or barrier) contribute to the continua (C$_1$ and C$_2$). 
This pathway arises because the higher energy pulse that is nominally exciting the barrier excitons is also resonant with the free carriers in the QWs and barrier, as represented in  Fig.~\ref{2Qpathways}. 
The rapid dephasing of this 2Q coherence is manifest in the extended peak along the E$_{2Q}$ axis. The 1Q coherence involving the free carriers also decays rapidly, as revealed in 1Q spectroscopy~\cite{Tollerud2016a}. In the H-L-B sequence the higher energy pulse arrives first, exciting this 1Q free carrier coherence. During the 300~fs before the next pulse arrives this coherence will decay, leaving only the 1Q barrier exciton coherence. The 2Q coherence that is generated then has no contribution from these free carriers and only the mixed 2-exciton coherence remains, as depicted in Fig.~\ref{2Qpathways}~(b). The detected signal in Fig.~\ref{SpectrumFull}(i) therefore only includes contributions from the mixed 2-exciton coherence. In contrast, the L-H-B sequence excites the QW exciton coherence with the first pulse, and the second pulse then generates double quantum coherences involving the barrier exciton and the free carriers, as depicted in Fig.~\ref{2Qpathways}~(a). Hence the signal from the H-L-B sequence shown in Fig.~\ref{SpectrumFull}(h) includes contributions from both CP$_{1(2)}$ and C$_{1(2)}$. It is important to note that the selectivity achieved in Fig.\ref{SpectrumFull}(i) requires both the spectral amplitude shaping and the delay between the first two pulses to filter out the free carrier contribution. See supporting information for a more thorough discussion of the available free carrier states and pathways.

Eliminating the signals involving EHP's in this manner -- and thereby isolating the mixed two-exciton signals -- allows us to extract additional information from the excitonic CPs, which can be incorporated as phenomenological constants in the optical Bloch equations to reproduce the salient features of the 2D spectra. Specifically, we can determine the precise position, shape, and linewidths of the CPs, which can provide useful information about dephasing and decoherence processes~\cite{Tollerud2016a}, separate excitonic and biexcitonic effects~\cite{Stone2009}, and identify many-body effects~\cite{Karaiskaj2010,Gellen}. Further details of different interactions and many-body effects can be determined from the real-valued 2Q 2D spectrum, which is included in the supporting information, as has been described previously for 2Q 2D data \cite{Karaiskaj2010} and extensively for 1Q rephasing and non-rephasing experiments~\cite{Li2006,Turner2012,Bristow2009,Nardin2014}.

The real valued spectra can also be useful to identify where positive and negative contributions from different nearby peaks overlap, leading to interference and peaks/dips in the absolute value spectra that can be misleading. Indeed, in the B-B-B pulse sequence the peaks of interest are convolved with other signal pathways, leading to interference that distorts the peak location, shape and amplitude. For example, the apparent peak positions in Fig.~\ref{SpectrumFull}~(g,h) place the mixed 2-exciton peak away from the expected location; in (i), however, with all other contributions removed it is apparent that it lies precicely at the energy matching the sum of the two excitonic transitions. On removing the EHP contribution we can also clearly see that the CPs in Fig.~\ref{SpectrumFull}~{(h)-(l)} have a tilted peak-shape, indicating that the E$_{2Q}$ width is limited by the sum of the single exciton linewidths, and not by carrier-carrier scattering~\cite{Tollerud2016a}. While an indication of the accurate peak locations can be obtained from the real-valued spectrum in the supporting information, the ability to isolate the relevant pathways provides unambiguous determination of peak position and shape.


The width along the E$_{2Q}$ direction of the signals involving the EHP indicates their short decoherence time, which in turn suggests that they could be suppressed in post-processing by windowing the data as a function of $t_{2Q}$ (as shown in supporting information). However, applying a filter in this time domain will affect the salient features in the 2Q 2D spectrum (such as peak shapes, widths and positions). Furthermore, such modification is very sensitive to the details of the windowing function -- particularly in feature dense spectra. In contrast, temporal filtering in $t_1$ using selective pulse sequences does not have the same confounding effect on the characteristics of double-quantum peaks because the full dynamics of the double quantum coherence are acquired and utilised.

In the case of the below-diagonal CPs in Fig.~\ref{SpectrumFull}~(k,l) the situation is a little more complicated. In the L-H-B sequence, the CP3 is clear, but it is also apparent that there is some contribution from free carriers, as expected for this pulse ordering from the preceding discussion. In this case, however, with emission from the barrier exciton, the continuum contribution is reduced and comes only from the 2Q coherence involving barrier exciton and the low energy QW continuum that can be excited by pulse-1.  In the H-L-B sequence, however, an additional contribution with emission from GaAs defect states becomes the dominant signal pathway.  This major change from L-H-B indicates that the pathway/s involving these defect states, which are excited by the higher energy pulse, become much more significant when the defects are excited by the first pulse compared to the case where the QW excitons are excited first. The absence of any contribution from this 2Q state with the L-H-B pulse sequence or in the above-diagonal CPs shows that the 2Q coherence responsible for this signal only arises when the initial excitation is to the defect state and the emission is from the defect state. This suggests that there is a doubly excited state that is only associated with the defect. Determination of the precise nature of this doubly excited state will be the subject of future work exploiting polarization control. Regardless, the point we make here is that the identification of this anomalous signal only becomes possible because of the spectral shaping and delays between the pulses generating the 2Q coherence.

\section{Summary}

We have demonstrated an experimental technique analogous to double-quantum heteronuclear NMR using pulse-shaper based double-quantum 2D electronic spectroscopy. In NMR these techniques are able to selectively identify specific atomic sequences. Here we demonstrate that controlling the pathway/s by which 2Q coherences are excited can significantly affect the measured 2D spectrum. This can allow specific peaks to be isolated from overlapping signals or hidden pathways to be identified. In the semiconductor quantum wells studied here,  signals from specific mixed two-exciton coherences are isolated by combining both spectral and temporal filtering to suppress the spectrally broad features created by rapidly dephasing free-carriers and defects. The mixed-exciton CPs that remain are unaffected, thereby enabling detailed analysis of the peak shapes and thus excitonic interactions. 
The combination of different pulse sequences also identifies an additional, unexpected pathway involving a doubly excited state associated with GaAs defects. The presence of this state only becomes evident when the pathway selective approach described here is used: from the broadband measurements the spectral signature would be confused for other interaction induced effects.

The capability to selectively excite specific two-exciton states in a deterministic order, as demonstrated here, can provide great understanding of excitonic interactions and many-body effects across a wide range of material systems~\cite{Chenu2014,Christensson2010,Richards2012,Naeem2015,Cassette2015,Mak2010,Wang2012b,Britnell2013}.  Further, the approach demonstrated overcomes several experimental challenges that have prevented optical analogues of established NMR techniques, and will enable future development of techniques that provide ever more incisive probes of multi-particle correlations and many-body effects in complex electronic systems.

\begin{acknowledgements}
This work was supported by the Australian Research Council.
\end{acknowledgements}

%
\bibliography{PS2Q,ExtraBib}

\end{document}